\documentstyle[aps,eqsecnum,preprint]{revtex} \baselineskip 12pt
\newcommand{\be}{\begin{equation}}
\newcommand{\ee}{\end{equation}}
\newcommand{\bea}{\begin{eqnarray}}
\newcommand{\eea}{\end{eqnarray}}
\begin{document}
\title{ PT invariant Non-Hermitian Potentials with Real QES Eigenvalues} 

\vspace{.4in}

\author{Avinash Khare }

\address{
Institute of Physics, Sachivalaya Marg,\\ Bhubaneswar-751005, India,\\
Email:  khare@iopb.res.in}

\author{ Bhabani Prasad Mandal}

\address{
Theory Group, Saha Institute of Nuclear Physics,\\
1/AF, Bidhannagar, Calcutta-700 064, India.\\
Email: bpm@tnp.saha.ernet.in
}

\vspace{.4in}

\maketitle

\vspace{.4in}

\begin{abstract}

We show that at least the quasi-exactly solvable eigenvalues of the Schr\"odinger equation with the
potential $V(x) = -(\zeta \cosh 2x -iM)^2$ as well as the periodic potential
$V(x) = (\zeta \cos 2\theta -iM)^2$
 are real for the PT-invariant non-Hermitian potentials in case the 
parameter $M$ is any odd integer. 
We further show that the norm as well as the weight functions for the
corresponding weak orthogonal polynomials are also real.
\end{abstract}

\newpage

In non-relativistic quantum mechanics, one usually chooses a real 
(Hermitian) potential so as to ensure the real energy eigenvalues of the
corresponding Schr\"odinger equation. About three
years ago, Bender and others \cite{bed,bed1,oth} have 
studied several complex potentials
which are invariant under the combined symmetry $PT$ and showed that 
even in all these cases, the energy eigenvalues of the Schr\"odinger equation
are real. This seems to suggest that instead of Hermiticity, it may be  
enough to have PT-invariant Hamiltonian so as to have real energy 
eigenvalues. It must however be noted that so far, this is merely a conjecture
which is being supported by several examples. Besides, even if this conjecture
is true, there are several basic questions which will have to be  
addressed before one can take the PT-invariant potentials more seriously.

In the absence of any rigorous proof, it is worthwhile studying new
PT-invariant non-Hermitian potentials and see if in each and every case
the eigenvalues are real or not.
The purpose of this note is to study
this question through two concrete examples.
In particular, we consider the system described by the
non-Hermitian PT-invariant Hamiltonian ($\hbar =2m =1$)
\begin{equation}
H= p^2 - (\zeta\cosh 2x -iM)^2 \, ,
\label{hh}
\end{equation}
where the parameter $\zeta $ is real and parameter $M$ has only
integer values. 
We show that at least the quasi-exactly solvable (QES) 
eigen values \cite{ush} of 
this Hamiltonian are real and bounded from below.
We also show that the corresponding set of polynomials form a family 
of weak orthogonal polynomials and further, the norms for these 
polynomials as well as the corresponding weight functions are real.
Using the duality transformation, we further show that the QES eigenvalues
of the complex, non-hermitian, but PT-invariant periodic potential
\be\label{2}
V(\theta) = (\zeta \cos 2\theta -iM)^2 \, , 
\ee
are also real.

Let us first show that the Hamiltonian (\ref{hh}) is 
PT-invariant. To that end, let us first notice that under $T$ 
reflection one replaces $i$ by $-i$ while under $P$ reflection one
replaces $x$ by $a-x$ where $a/2$ is the origin about which one is
performing the parity reflection. It is then easily checked that the
Hamiltonian (\ref{hh}) is indeed $PT$ symmetric under
the parity reflection $x \rightarrow i\pi /2 -x$.

We substitute
\begin{equation}
\psi(x) = e^{i\frac{ \zeta}{2}\cosh 2x}\phi(x) \, ,
\end{equation}
in the Schro$\ddot{o}$dinger equation $H\psi= E\psi$ with $H$ as given by
Eq. (\ref{hh}) and obtain
\begin{equation}
\phi^{\prime\prime}(x) +2i\zeta\sinh 2x\phi^\prime (x) + \left
 [(E-M^2+\zeta^2) -2i(M-1)\zeta\cosh 2x \right ]\phi(x) =0 \, .
\label{1st} 
\end{equation}
Ordinarily, the boundary conditions that give quantized energy levels are
$\psi (x) \rightarrow 0$ as $\mid x \mid \rightarrow \infty$ on the real axis.
However, in the present 
case, we have to continue the eigenvalue problem into the 
complex-x plane \cite{bt}. 
On putting, $x = u+iv$ where $u,v$ are real, it is easy to 
see that for $u > 0$ the above boundary condition is satisfied so long as 
$ -\pi < v < -\pi/2 \ (mod\ \pi$) while for $u < 0$ it is satisfied if
$-\pi/2 < v <0 \ (mod \ \pi$). 

On further substituting
\begin{equation}
z=\cosh 2x -1;\ \ \ 
\phi = z^s\sum_{n=0}^\infty \frac{ R_n(E)}{n!} 
\left ( \frac{ z+2}{2} \right )^{ \frac{ n}{2}} \, ,
\label{2nd}
\end{equation}
we obtain the three-term recursion relation $(n\ge 0)$
\begin{eqnarray} 
R_{n+2}(E)&-&\left [ n^2 +4(s-i\zeta)n + 4s^2 + \left .
E-M^2+\zeta^2+2i(M-1)\zeta \right .  \right ]R_n(E) \nonumber \\  
-&&4\zeta \left [ M+1-2s-n \right ]n(n-1) R_{n-2}(E)=0 \, ,
\label{r1}
\end{eqnarray}
provided $2s^2 = s$ i.e. either $s=0$ or
$s= \frac{ 1}{2}$. Thus we have two sets of independent solutions; one for
$s=0$ and other for $s= \frac{ 1}{2}$. Note that the parameter $s$ is not
contained in the potential and this is perhaps related to the fact that 
for any odd M $(>1)$, the QES solutions corresponding to both even and odd
number of nodes are obtained.  

>From Eq. (\ref{r1}) we observe that the even and odd
polynomials $R_n(E)$ do not mix with each other and hence 
we have two separate three-term recursion relations
depending on whether $n$ is odd or even. In particular, it is easily 
shown that the three-term recursion relations corresponding to the even and odd 
$n$ cases respectively are given by $(n\ge 1)$
\begin{eqnarray}
P_n(E)&-& \left [ 4n^2 +8n(s-i\zeta-1)+4s^2-8s +4+6i\zeta +E -(M-i\zeta)^2
\right ]P_{n-1}(E) \nonumber \\   
&+& 8i\zeta(n-1)(2n-3)\left ( M+3-2s-2n \right ) P_{n-2} = 0 \, ,
\label{r2}  \\  
Q_n(E)&-& \left [ 4n^2 +4n(2s-2i\zeta-1)+4s^2-4s +1+2i\zeta +E
-(M-i\zeta)^2
\right ]Q_{n-1}(E) \nonumber \\  
&+& 8i\zeta(n-1)(2n-1)\left ( M+2-2s-2n \right ) Q_{n-2} = 0  \, , 
\label{rr1}
\end{eqnarray}
with $P_0(E)=1, Q_0(E)=1. $

We find that for odd integer values of $M$, there are $M$ quasi-exactly
solvable eigen values and all of them are real provided $|\zeta|\le \zeta_{c}$,
where $\zeta_{c}$ is a function of $M$.
At $\zeta = \zeta_{c}$
two highest eigen values become degenerate, while for $\zeta
>\zeta_{c}$ some of the eigen values are complex. In particular, if $M = 2k+1$
($k=0,1,...$), then $k+1$ and $k$ levels with real energy eigenvalues 
are obtained from the zeros of the
orthogonal polynomials $P_{k+1} (E)$ and $Q_k (E)$ respectively which we term
as the critical polynomials. 
Further, all higher $P(Q)$ polynomials exhibit factorization property i.e.
\bea\label{3}
P_{k+n+1} (E) & = & P_{k+1} (E) {\bar P}_{n} (E) \nonumber \\
  Q_{k+n} (E) & = & Q_k (E) {\bar Q}_n (E) \, ,
\eea  
where ${\bar P}_0 (E) =1 = {\bar Q}_0 (E)$.
It is interesting to note that even though the 
polynomials satisfying recursion relations 
(\ref{r2}) and (\ref{rr1}) are complex, all critical polynomials (i.e. 
$P_{k+1}$ and $Q_k$) are real in case $M = 2k+1$. 

We now consider several odd integer values of $M$ and show explicitly
that the QES eigen values are real for $\zeta\le \zeta_{c}$. 
Let us start with $M=1$, in which case 
$P_1(E) \equiv E-1+\zeta^2$ is the critical
polynomial and the only QES level is $ E =1-\zeta^2$ which is always real.
It is not very clear if this is the ground state energy of the system or
not even though for the corresponding real potential, for M=1 one indeed
obtains the ground state energy \cite{qes2}.

For $M=3$ the critical polynomials are,
\begin{eqnarray} 
P_2 &=& {\cal E}^2 +4{\cal E} +16\zeta^2 \nonumber \\ 
Q_1 &=& {\cal E} +4 \, ,
\label{m3}
\end{eqnarray}
where 
\be\label{e}
{\cal E} \equiv E-M^2 +\zeta^2 \, .
\ee
The energy levels are
\begin{eqnarray} 
E &=& 5- \zeta^2 \nonumber \\ 
E &=& 7- \zeta^2-2\sqrt{1-4\zeta^2} \nonumber \\ 
E &=& 7- \zeta^2+2\sqrt{1-4\zeta^2}  
\label{e3}
\end{eqnarray} 
of which the first energy is obtained from $Q_1=0$ while the other two 
are obtained from $P_2 =0$.
For $\zeta^2\le \zeta_{c}^2 = 1/4$ these levels are real
and at $\zeta =\zeta_{c}$ highest two levels are degenerate.

Proceeding in the same way, we have considered the cases of $M=5,7,9$
but unfortunately in these cases (except when M=5 and $Q_2=0$) 
one has to solve cubic or higher order
equations so as to obtain the corresponding energy eigenvalues. We have
calculated these eigenvalues in case $\zeta^2 = 0.01, 0.02$ and $0.025$ 
(all of
which are less than the corresponding $\zeta^2_{c}$ and they are given in
Tables I, II and III. In each case, we also give the expressions for the 
corresponding critical polynomials as well as $\zeta_c$. 

We have studied the properties of the weakly orthogonal polynomials 
$P_n (E)$ and $Q_n (E)$ and we find that many of these are 
similar to the Bender-Dunne polynomials \cite{bd}. In particular, since in
both the cases 
 the recursion relations are similar to those given by  Eq. (\ref{r2}),
hence for
all values of the parameters $M$ and $s$, they form an orthogonal set.
Secondly, the wave function $\psi(x,E)$ is the generating function
for the
polynomials $P_n(E)$ as well as $Q_n(E)$. Thirdly when $M$ is a
positive
integer, both of these polynomials  exhibit factorization property
whose precise form
depends on whether $M $ is even or odd. The norms and weight functions for
these polynomials can be calculated using the same procedure as described
in \cite{qes2} and they turn out to be complex. 

We now show that 
there is an alternative way \cite{fin} to express this system as an
algebraic quasi-exactly solvable system and that in this case
the norms and the weight
functions of the corresponding orthogonal polynomials are {\it real}.
 
Following Finkel et al. 
\cite{fin}, we change the variable, $ z= e^{2x} $ and make 
the gauge transformation
\begin{equation}
\mu (z) = z^{ \frac{ 1-M}{2}}e^{i \frac{ \zeta}{4}(z+ \frac{ 1}{z})} \, ,
\label{sub}
\end{equation}
in the Hamiltonian (\ref{hh})
to map it to a differential operator, $H_g$ ( called as gauge Hamiltonian).
This gauged Hamiltonian, $H_g$ can be expressed in terms of the generators
of Sl(2,R) by 
\begin{equation}
H_g(z) = -4J_0^2 -2i\zeta J_{+} +2i\zeta J_{-} - c^* \, ,
\label{hg}
\end{equation}
where $c^* = -M^2 +\zeta^2 $ while the generators $J_{+,-,0}$ are given by
\be\label{gg}
J_{-} = \frac{\partial}{\partial z} \, , \ J_{O} = z\frac{\partial}{\partial z}
-\frac{n}{2} \, , \ J_{-} = z^2 \frac{\partial}{\partial z} -nz \, . 
\ee
Now on substituting
\begin{equation}
\phi (z) = \sum_{n=0}^{\infty} \frac{ R_n(E)}{(2i\zeta)^n n!} z^n \, ,
\end{equation}
in the Schr$\ddot{o}$dinger equation, $H_g\phi = E\phi $, one finds that 
$R_n (E)$ satisfy a three-term recursion relation given by 
\begin{equation}
R_{n+1}(E) = (E-b_n)R_n(E) -a_nR_{n-1}(E), \ \ \ \ \ \ n\geq 0 \, ,
\label{arr}
\end{equation}
where
\begin{eqnarray}
a_n &=& - 4n(M-n)\zeta^2 \, , \nonumber \\
b_n &=& 4n(M-1-n) + 2M - 1 -\zeta^2 \, .
\label{akbk} 
\end{eqnarray}
Thus we see that (with $R_{0}=1$) the same system (\ref{hh}) forms 
another family of weakly
orthogonal, monic polynomials.
The first few polynomials generated by the recursion relation (\ref{arr})
are given by
\begin{eqnarray}
R_1 &=& E-b_0 \nonumber \\ 
R_2 &=& E^2 -(b_0+b_1)E + a_1 +b_1b_0 \nonumber \\ 
R_3 &=& 
E^3 -(b_0+b_1+b_2) E^2 + ( b_0b_1+b_0b_2+b_1b_2+a_1
+a_2)E \nonumber \\
    &-& (b_0b_1b_2+a_1b_2+a_2 b_0) \nonumber \\ 
R_4 &=& (E-b_3)R_3 - a_3 R_2 \, , 
\label{aps}
\end{eqnarray}
where $a_n$ and $b_n$ are given by the Eq. (\ref{akbk}). 

These polynomials are also of the Bender-Dunne type and exhibit
similar properties as the Bender-Dunne polynomials. For $M=$ positive
integer (say $k$,) these polynomials exhibit factorization property
given by,
 \begin{equation}
R_{k+n} = R_k \bar{R}_{n} \ \ \ \  n\ge 0 \, .
\label{fac}
\end{equation}
Note that at $M=k$, $a_k = 0$ and hence Eq. (\ref{arr}) reduces to two term 
recursion relation. 
Further, when M is odd (say M=2k+1) then not surprisingly we find that
the corresponding critical polynomial factorizes as the product of $P_{k+1}$
and $Q_{k}$, i.e. 
\be\label{10}
R_{2k+1} (E) = P_{k+1} (E) Q_{k} (E) \, ,
\ee
where $P_k,Q_k$ are as defined by Eqs. (\ref{r2}) and (\ref{rr1}).

Here the polynomials $\bar{R}_n$ correspond to the non-QES part
of the spectrum and they satisfy the three-term recursion relation 
\begin{equation}
\bar{R}_{n+1} = ( E-b_{M+n})\bar{R}_n -a_{M+n}\bar{R}_{n-1}
\end{equation}
Hence these polynomials (with $\bar{R}_0 =1$) 
also form a set weakly orthogonal polynomials.

The norms (squared) of the orthogonal set of polynomials $R_n (E)$ 
can be easily determined using the recursion relations of these
polynomials and one finds
\begin{equation}
\gamma _n = (-1)^n\prod_{i=1}^n a_i \, ,
\end{equation}
where $a_i$ are given by eq. (\ref{akbk}). 
It is clear that the norms for these polynomials, even though real, 
are not always positive. The
norms for the even polynomials are positive  and on the other hand the
norms for the odd polynomials are negative so long as $n <M$ while 
for $n \ge M$,
$\gamma_n$ vanishes. Similarly, the weight functions, $\omega
(E) $ for these polynomials are also real, though not positive definite.   
This is easily seen using the proposition 2.1 in \cite{fin1} and
the fact that $a_k $ are negative for $1\le k\le M$.

Before finishing this note, we show that the QES eigenvalues of a related
complex non-Hermitian but PT-invariant periodic potential are also real. 
To that end,
it is worth noting that
recently Krajewska et. al. \cite{kra} have discussed
the consequences of the
anti-isospectral transformation (also termed as duality transformation)
in the  QES problems. In particular, they have shown that under the
transformations $x\rightarrow  ix =y $, if a potential 
$V(x)$ goes to $\bar{V}(y)$
then  the QES levels of the two  are also related. In particular, they have 
shown that if $M$ levels of the potential $V(x)$ are QES levels  with
energy eigenvalues and eigenfunctions $E_k(k=0,1, \cdots, M-1)$ and $\psi_k(x)$
respectively then the energy eigenfunctions of $\bar{V}(y)$ are given by
\begin{equation}
\bar{E}_k = -E_{M-1-k}, \ \ \ \bar{\psi}_k(y) = \psi_{M-1-k}(ix)
\label{35}
\end{equation}

We shall now apply this anti-isospectral transformation to the system described
by Eq. (\ref{hh}) and obtain another example of a non-Hermitian but 
PT-invariant system with real QES energy eigenvalues.
 On applying the duality transformation $x\rightarrow i \theta $ to the
Schro$\ddot{o}$dinger equation $ H\psi = E\psi $ with $H$ as given by
Eq. (\ref{hh}) we obtain the following Schr$\ddot{o}$dinger equation for
the PT-invariant DSG potential 
\be\label{36}
\left [-\frac{ d^2}{d \theta ^2}+\left (\zeta\cos2 \theta -iM \right )^2
\right ]\psi(\theta )= \hat{E}\psi( \theta ), \ \ \  \hat{E}= -E.
\ee

Using Eq. (\ref{35}), the QES levels for this system can be obtained for odd
integer values of $M$  from the QES levels of the dual system which we have
already discussed in details. For example, for $M=1$, the only QES level
is,
\begin{equation}
\hat{E}= -(1-\zeta^2) \, ,  \ \ \ \ \ \ \ \psi = e^{+i\frac{ \zeta}{2}
\cos 2 \theta } \, ,
\end{equation}
which is real for any value of $\zeta$ . On the other hand for $M=3$ the three
QES levels are given by
\begin{eqnarray}
\hat{E} &=& -7+\zeta^2-2\sqrt{1-4\zeta^2} \, ,  \ \   
\psi= \left [-2i\zeta-
\left \{ \sqrt{1-4\zeta^2}-1 \right \}\cos 2 \theta \right ]e^{i\frac{\zeta}{2}
\cos 2 \theta } \, , \nonumber \\
\hat{E} &=&  \zeta^2 -5 \, ,  \ \ \psi=\sin 2 \theta e^{i\frac{ \zeta}{2}
\cos 2 \theta } \nonumber \\   
\hat{E} &=& -7+\zeta^2+2\sqrt{1-4\zeta^2} \, ,  \ \  \psi= \left [-2i\zeta+
\left \{ \sqrt{1-4\zeta^2}+1 \right \}\cos 2 \theta \right ]e^{i\frac{\zeta}{2}
\cos 2 \theta } \, . 
\label{123} 
\end{eqnarray}
Similarly one can easily obtain the QES levels for the cases, $M=5,7, 9
\cdots $ etc by replacing $E$ by $-E$ in the corresponding expressions
for the complex DSHG case.  

One can now study the Bender-Dunne polynomials of this dual system and it
is easy to see that most of the discussion of the previous system 
goes through in this case.

When $M$ is even integer then the eigenvalues are complex conjugate
pair. Further, one can show that whereas PT symmetry remains unbroken
when $M$ is odd integer, it is spontaneously broken when $M$ is even
integer \cite{new}.

Summarizing, we have seen that at least the QES eigenvalues of the 
non-Hermitian PT-invariant potentials (\ref{hh}) and (\ref{36}) 
are real and bounded from below. It would be nice if one can also compute 
the non-QES spectrum of these potentials and show that even these energy
eigenvalues are real. 

{\bf Acknowledgment:} We are grateful to Prof. Carl M. Bender for 
pointing out the PT-invariance of the potentials discussed in this
paper.

\newpage

{\bf Table I : QES levels  for the case  M=5 [$\zeta^2_c \simeq .08757$]
 }

\vspace{.5in}

\begin{center}

\begin{tabular}{|c|c|c|c|}\hline
Energy Levels & $\zeta^2 =.01$ & $\zeta^2 =.02 $& $\zeta^2=.025$ \\
&&& \\ \hline
$E_P$ & \ \ \ 9.00331480\ \ \ & \ \ \ 9.00659241\ \ \ &\ \ \ 9.00821724
\ \  \\ &&& \\ \hline
$E_Q$ & 9.00334818  & 9.00672620 & 9.00842644 \\ &&& \\ \hline
$E_Q$ & 20.97665182  & 20.95327381 & 20.94157356 \\ &&& \\  \hline
$E_P$ & 21.10018724  & 21.20836270  & 21.26599253 \\ &&&  \\ \hline
$E_P$ & 24.86649797 & 24.72504489 & 24.65079023\\ &&& \\ \hline
\end{tabular}

\end{center}

The critical polynomials in this case  are given by,

\begin{eqnarray}
P_3 &=& {\cal E} ^3 +20 {\cal E} ^2 +{\cal E} (64 +64\zeta^2)
+768\zeta^2 \nonumber \\
Q_2 &=& {\cal E} ^2 +20 {\cal E} +64 +16\zeta^2
\label{m5}
\end{eqnarray}

\newpage

{\bf Table II :QES levels for the case  M=7 [$\zeta_c^2 \simeq .04435$]  }

\vspace{.5in}
\begin{center}

\begin{tabular}{|c|c|c|c|}\hline
Energy Levels & $\zeta^2 =.01$ & $\zeta^2 =.02 $ &$ \zeta^2 =.025 $ \\
&&& \\ \hline
$E_Q$ & \ \ \ 13.00199970 \ \ \ &\ \ \ 13.00399883 \ \ \ & \ \ \ 13.00499708 \ \
\ \\ &&& \\ \hline
$E_P$ & 13.00199971 & 13.00399878  & 13.00499918 \\ &&& \\ \hline
$E_P$ & 33.01123852 & 33.02228660 & 33.02773885 \\ &&& \\ \hline
$E_Q$ & 33.01140573 & 33.02295762 & 33.02878902 \\ &&& \\ \hline
$E_Q$ & 44.95659457 & 44.91304361 & 44.89121290 \\ &&& \\ \hline
$E_P$ & 45.21134926 & 45.46170007 & 45.60896931 \\ &&& \\ \hline
$E_P$ & 48.73541251 & 48.43201450 & 48.25829367\\ &&& \\ \hline
\end{tabular}

\end{center}

The critical polynomials for this case are,

\begin{eqnarray}
P_4 &=& {\cal E}^4 +56 {\cal E} ^3 +(784+160\zeta^2) {\cal E} ^2 +
(2304  + 6528\zeta^2){\cal E} \nonumber \\  &+& 2304 \zeta^4 +55296\zeta^2
\nonumber \\
Q_3 &=& {\cal E}^3 +56 {\cal E}^2 +(784+64\zeta^2) {\cal E}
+2304+1536 \zeta^2
\label{m7}
\end{eqnarray}

\newpage

{\bf Table III : QES levels for the case M=9 [$\zeta_c^2\simeq .02675]$ }

\vspace{.3in}

\begin{center}

\begin{tabular}{|c|c|c|c|}\hline
Energy Levels & $\zeta^2 =.01$ & $\zeta^2 =.02 $& $\zeta^2 =.025 $\\
&&& \\ \hline
$E_P$ & \ \ \ 17.00096869 \ \ \ & \ \ \ 17.00182564 \ \ \ & \ \ \ 
17.00225406 \ \ \ \\  &&& \\ \hline
$E_Q$ & 17.00153729 & 17.00295705 & 17.00366471 \\ &&& \\ \hline
$E_Q$ & 45.00607980 & 45.01268446 & 45.01599526 \\ &&& \\ \hline
$E_P$ & 45.00751350 & 45.01552516 & 45.01952952 \\ &&& \\ \hline
$E_P$ & 65.02149678  & 65.04164257 & 65.05150021 \\ &&& \\ \hline
$E_Q$ & 65.02298207 & 65.04558083 & 65.05703252 \\ &&& \\ \hline
$E_Q$ & 76.92940084 & 76.85877767 & 76.8233075 \\ &&& \\ \hline
$E_P$ & 77.37417833  & 77.91167617 & 78.38610328 \\ &&& \\ \hline
$E_P$ & 80.5458427 & 79.92933046 & 79.41561294 \\ &&& \\ \hline
\end{tabular}

\end{center}

The critical polynomials for this case are,

\begin{eqnarray}
P_5 &=& {\cal E}^5 +120 {\cal E}^4 +(4368+320\zeta^2) {\cal E}^3
+(52480  +30280 \zeta^2) {\cal E}^2 \nonumber \\  &+& (147465 +806912
\zeta^2+16384 \zeta^4) {\cal E} +655360 \zeta^4 + 5898240 \zeta^2
\nonumber \\
Q_4 &=& {\cal E}^4+120 {\cal E}^3+ (4368+160\zeta^2) {\cal E}^2
+(52480 +11648\zeta^2){\cal E} \nonumber \\
&+& 147456 +182272\zeta^2
\label{m9}
\end{eqnarray}


\newpage
\begin{thebibliography}{99}
\bibitem{bed} C. M. Bender and S. Boettcher, Phys. Rev. lett. {\bf 80} 5243
(1998).
\bibitem{bed1} C. M. Bender and S. Boettcher, J. Phys. {\bf A31} (1998)
L273. 
\bibitem{oth} C. Bender , S. Boettcher, and P. Meisinger, J. Math. Phys.
{\bf 40} 2210, (1999); 
C. M. bender, S. Boettcher, H. F. Jones and
Van M. Savage, quant-ph/9906057; C. M. Bender and
G. V. Dunne, J. Math. Phys. {\bf 40}
4616,(1999); C. M. Bender, G. V. Dunne and P. Meisinger,
Phys. Lett {\bf A252}, 272, (1999); Bender, Dunne and
Meisinger J. Math. Phys {\bf 40 } 2201, (1999); Bender and K. A. Milton ,
hep-th/9802184; F. M. Fernandez et. al., 
Quant-ph 9812026.
\bibitem{ush} A. Ushveridze, {\it Quasi-Exactly Solvable Models
in Quantum Mechanics}, Inst. of Physics Publishing, Bristol, (1994), and
references their in. 
\bibitem{bt} C.M. Bender and A. Turbiner, Phys. Lett. {\bf A173} (1993) 442.  
\bibitem{qes2} A. Khare and B. P. Mandal,  J. Math. Phys. {\bf 39}, 3476,
(1998).
\bibitem{bd} C. M. Bender and G. V. Dunne, J. Math. Phys. {\bf 37} (1996) 6.
\bibitem{fin} F. Finkel, A . Gonzaler-lopez and M.A. Rodriguez, 
J. Math. Phys. {\bf 40} (1999) .
\bibitem{fin1}  F. Finkel, A. Gonzalez-Lopez And M. A. Rodriguez,
J. Math. Phys. {\bf 37} (1996) 3954.
\bibitem{kra} A. Krajewska, A. Ushveridze and Z. Walczak,
Mod. Phys. Lett. {\bf A 12} (1997) 1225.
\bibitem{new} A. Khare and B. P. Mandal, Phys. Lett A (In press)
 quant-ph/0006126 
 
\end{thebibliography}
\end{document}